\documentclass[reqno ,11pt]{amsart}

\usepackage[foot]{amsaddr}
\usepackage{graphicx}
\usepackage[usenames,dvipsnames]{xcolor}
\usepackage[paperwidth=7in,paperheight=10in,text={5in,8in},left=1in,top=1in,headheight=0.25in,headsep=0.4in,footskip=0.4in]{geometry}
\usepackage{subfigure}
\usepackage{lineno}
\usepackage{hyperref}

\definecolor{linenocolor}{gray}{0.6}
\definecolor{prec}{RGB}{42,115,205}
\definecolor{sens}{RGB}{255,112,0}
\definecolor{spec}{gray}{0.4}

\usepackage{fix-cm}

\newcommand{\mr}{\mathrm}

\usepackage{mathpazo}
\makeatletter
\renewcommand{\@biblabel}[1]{\quad#1.}
\makeatother

\usepackage{cite}

\begin{document}

\title[The Population Dynamics of Scientific Discovery]{\large Replication, Communication, and the Population Dynamics of Scientific Discovery}
\author[McElreath \& Smaldino]{Richard McElreath$^{1,2}$ \and Paul E.~Smaldino$^1$}
\address{$^1$Department of Anthropology, UC Davis, One Shields Avenue, Davis CA 95616}
\address{$^2$Center for Population Biology, UC Davis}
\email{mcelreath@ucdavis.edu}


\maketitle

{\vspace{-6pt}\footnotesize\begin{center}\today\end{center}\vspace{24pt}}


\noindent \textbf {Many published research results are false \cite{ioannidis_why_2005}, and controversy continues over the roles of replication and publication policy in improving the reliability of research. Addressing these problems is frustrated by the lack of a formal framework that jointly represents hypothesis formation, replication, publication bias, and variation in research quality. 
We develop a mathematical model of scientific discovery that combines all of these elements. This model provides both a dynamic model of research as well as a formal framework for reasoning about the normative structure of science. We show that replication may serve as a ratchet that gradually separates true hypotheses from false, but the same factors that make initial findings unreliable also make replications unreliable. The most important factors in improving the reliability of research are the rate of false positives and the base rate of true hypotheses, and we offer suggestions for addressing each. Our results also bring clarity to verbal debates about the communication of research. Surprisingly, publication bias is not always an obstacle, but instead may have positive impacts---suppression of negative novel findings is often beneficial. We also find that communication of negative replications may aid true discovery even when attempts to replicate have diminished power. 
The model speaks constructively to ongoing debates about the design and conduct of science, focusing analysis and discussion on precise, internally consistent models, as well as highlighting the importance of population dynamics. }

\vspace{12pt}



\newpage

\section*{Introduction}
Imagine two of your close colleagues have just heard about attempts to replicate their positive research findings. Colleague A is thrilled that the attempt was successful. Colleague B is upset that the attempt was unsuccessful. What is the probability that Colleague A's hypothesis is true? What is the probability that Colleague B's hypothesis  is false? 

This is not a fair quiz, because in truth no one knows the answers to these questions. The absence of replication in many fields \cite{makel_replications_2012,franco_publication_2014,schmidt_shall_2009}, combined with the absence of a formal framework for understanding replication, makes it difficult to even outline an answer. 
In the absence of replication, there is substantial concern that many published findings may be false \cite{ioannidis_why_2005}, an argument with empirical support \cite{begley_drug_2012,prinz_believe_2011,sullivan_spurious_2007}. The history of science buttresses these observations. A recent catalog of false discoveries of chemical elements outnumbers the current number of real elements in the periodic table \cite{Fontani:2014aa}. 
In addition to concerns about replication are concerns about research practice and publication bias. Without knowing how many studies were conducted but not published, it is not possible to assign evidential value to either initial findings or replications. And it is not yet easy to acquire empirical evidence about these factors, as even the best empirical studies of publication bias still rely upon researcher self-report \cite{franco_publication_2014}.

Thus many opinions can be sustained about the evidential value of both initial findings and replications. 
As a result, recent controversies over failed replications demonstrate a lack of consensus on norms for replication and publication \cite{bissell_reproducibility_2013,bohannon_replication_2014,kahneman_new_2014,schnall_clean_2014}. What is the evidential value of replication, positive or negative? What is the impact of publication bias \cite{rosenthal_file_1979}?  If replication is part of an ``invisible hand''\cite{hull_science_1988} that corrects scientific errors, how much replication is needed? And what are the risks of poorly designed or interpreted replication attempts \cite{bissell_reproducibility_2013}? When replication is not possible or practical, what other measures can be taken to improve the reliability of research?

These questions remind us that little is understood about the population dynamics of discovery, replication, and scientific communication. Much more attention has been given to individual methods of research design and data analysis.  
And while it is useful to analyze research methods in isolation, such calculations are unsatisfying. A lot of research activity is hidden from the public record. This means the actual number of findings for an hypothesis may never be known \cite{rosenthal_file_1979}. And since researchers select hypotheses for further study from the literature itself, findings and publication biases cascade into other findings, interacting with biases and incentives \cite{orourke_meta-analysis_1989}. 

To know the evidential value of research, we must study the population dynamics that produce it \cite{campbell_toward_1985,hull_science_1988,popper_conjectures_1963,Kitcher2000}. 
So here we construct and solve a mathematical model of scientific beliefs formed by a population of boundedly rational agents who accumulate evidence for and against hypotheses. We adopt a general signal detection framework that may apply to diverse statistical paradigms, whether $p$-valued or Bayesian. We study the joint dynamics that arise from replication, publication bias, and differences in research quality between original studies and replications.   Our goal is not to accurately simulate science, but rather to understand it better using the same reductionist tools that have been so successful in illuminating population dynamics more generally \cite{Levins:1966fk,Wimsatt:1987aa}. Our model implicitly provides, for example, a neutral model of scientific dynamics in which all hypotheses are false and yet discoveries are continuously published. It also provides a range of ``selectionist'' models that might be compared to data. The clarity of a quantitative framework will stimulate and clarify the development of later empirical investigation and experimental intervention.

The paper proceeds by first outlining the dynamic structure of the model. 
We then solve the model for both its long-run dynamics and its epistemological implications---what should a rational agent believe about an hypothesis, given a record of published results? We present a general interpretation of the joint dynamics, so the reader can extrapolate lessons from our simple model to the complexity and diversity of real science. We conclude by relating our results to ongoing debates about improving the reliability of scientific research.

\section*{Model Description}

\begin{figure}[tp]
\begin{center}
	\hspace{-0.5in}\includegraphics[width=5.5in]{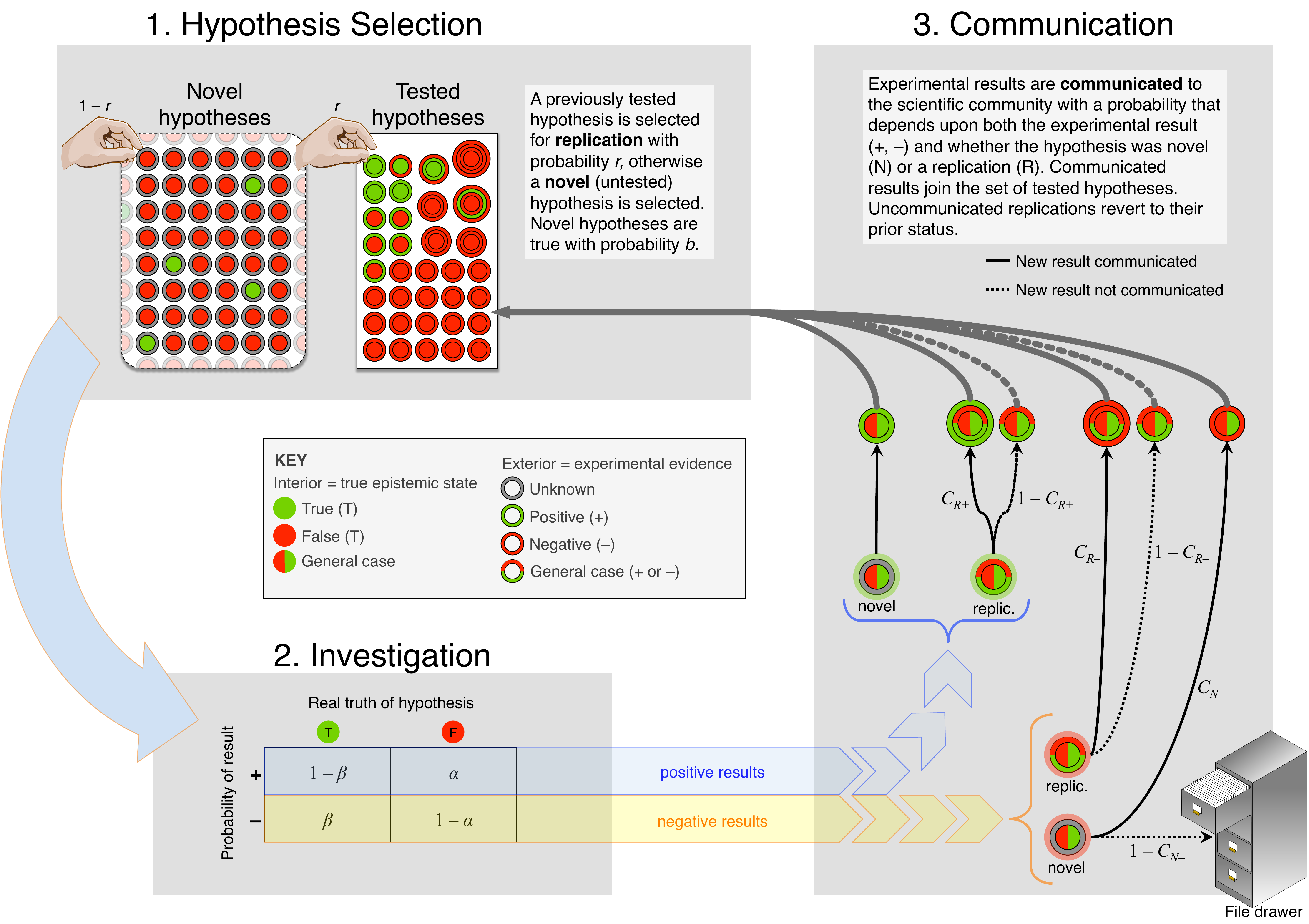}
\caption{\small Population dynamics of replication.}
\label{fig_flowchart}
\end{center}
\end{figure}

The model is illustrated in Fig.~\ref{fig_flowchart}. We have also constructed an interactive, web-based tutorial on the conceptual foundations of the model, as well as fully adjustable simulation code, available at {\url http://xcelab.net/replication/}. 
A population of researchers studies many different hypotheses. Each hypothesis is either \emph{true} (green) or \emph{false} (red). These hypotheses could be simple associations, such as \emph{green jelly beans cause acne} \cite{Munroe882}, or more general claims, such as \emph{evolution is predictable}. Research results in either a \emph{positive} or a \emph{negative} finding. These findings may be the result of formal hypothesis tests or informal assessments. True hypotheses produce positive findings more often than do false hypotheses, but the researchers never know for sure which hypotheses are true. Under these assumptions, the only information relevant for judging the truth of an hypothesis is its \emph{tally}, the difference between the number of published positive findings and the number of published negative findings for each hypothesis, and we summarize results in terms of these tallies. In reality, much other information is relevant to judging the truth of an hypothesis. Our assumptions are tactical ones. More complex models of scientific communication are possible, but any such model must include the components in our model, and so our results establish a critical baseline. 

Each time interval, research activity has three stages that alter these tallies. In stage 1 (Fig.~\ref{fig_flowchart}, upper-left) each researcher chooses to investigate  one of $n$ previously published hypotheses, with probability $r$, or a novel hypothesis, with probability $1-r$. When replicating, a researcher chooses a previously published hypothesis at random and performs a new study of it. Later, we allow researchers to target hypotheses with specific tally values, rather than choosing at random. A novel hypothesis is true with probability $b$, the \emph{base rate}, reflecting mechanisms of hypothesis formation. Untutored intuition, for example, may be expected to yield a very low $b$. Genome wide association studies likewise have low $b$, because relatively few loci are associated with any particular phenotype. There is no consensus on base rate, except that most scientists we know believe their own personal $b$ values are better than average. So we allow $b$ to vary freely in the model.

In stage 2, a true hypothesis produces a positive finding $1-\beta$ of the time, its \emph{power}. A false hypothesis produces a positive finding $\alpha$ of the time, its \emph{false positive rate}. We assume that $1-\beta>\alpha$. Later we allow the values of $\beta$ and $\alpha$ to differ between replication attempts and initial studies. Note that $\beta$ and $\alpha$ are not merely properties of a statistical procedure, but rather of an entire investigation. For example, using several procedures and selecting the one that produces a positive result will inflate $\alpha$ \cite{simmons_false-positive_2011-1}.

In stage 3, findings may be communicated to other researchers. Not every finding is communicated, either because no one tries to communicate it or rather because it cannot be published. Only communicated findings can adjust a tally. Let $c_{\mathrm N-}$ be the probability that a negative ($-$) finding about a new (N) hypothesis is communicated. We assume for simplicity that all new positive results are communicated ($c_{\mathrm N+}=1$). Even though replication findings are evidentially equivalent to novel findings, they may be communicated with different probability. Let $c_{\mathrm R-}$ and $c_{\mathrm R+}$ be the probabilities that replications with negative and positive findings, respectively, are communicated. 

These assumptions define the dynamics of the expected numbers of true and false hypotheses with a given tally. We present the full recursions in  the Supporting Material. In the simplest case (full communication: $c_{\mathrm N-}=c_{\mathrm R-}=c_{\mathrm R+}=1$), the number $n_{\mathrm T,s}$ of true hypotheses with an observed tally $s$ in the next time step is given by:
\begin{equation}
	n_{\mathrm T,s}^\prime = n_{\mathrm T,s} + 
	a n r \left( 
	-\frac{n_{\mathrm T,s}}{n} + \frac{n_{\mathrm T,s-1}}{n}(1-\beta) +  \frac{n_{\mathrm T,s+1}}{n}\beta \right)
\end{equation}
where $a>0$ is the rate of research activity as a proportion of $n$. This expression says that the number in the next time step is just the current number plus all of the flows in and out caused by replications. 
In the case that $s=-1$ or $s=1$, there is an additional term $an(1-r)b\beta$ or $an(1-r)b(1-\beta)$, respectively, to represent the inflow of novel findings. 
Recursions $n_{\mathrm F,s}^\prime$ for false hypotheses are constructed from a change in variables: $1-\beta \rightarrow \alpha$, $b \rightarrow 1-b$. 
Notice that this implies that the model is easily extended to any number of hypothesis types, such as effect size differences, that differ in power and false-positive rate. We analyze the \emph{true}/\emph{false} dichotomy because of its prominence and simplicity.

\section*{Analysis} 

By literature review, a tally can be constructed for any given hypothesis. Given an observed tally, but a number of possibly unobserved studies, what is the probability that an hypothesis is correct? The model allows us to address this question for a diversity of scenarios. Before presenting the solutions, note that the answers that the model provides can be understood both from a pure population dynamics perspective and from a probabilistic reasoning perspective. From the dynamics perspective, the population will converge from any initial condition to a unique steady state in which the solutions give \emph{frequencies} of true hypotheses at each tally value. Equally valid is the epistemological perspective that the solutions tell us for any unique hypothesis the \emph{probability} it is true, given a state of information \cite{Cox:1946aa}. One consequence of this is that the solutions do not require that all hypotheses share the same parameter values.

For each tally value $s$, we solved for the steady state proportions of true and false hypotheses, $\hat p_{\mathrm T,s}$ and $\hat p_{\mathrm F,s}$. We also derived the same solutions under the probabilistic interpretation, and verified our solutions numerically and through stochastic simulation. We present complete analytical solutions in  the Supporting Material. In the simplest case (for full communication), solutions take the form:
\begin{align}
	\hat p_{\mathrm T,s} = b(1-r) \sum_{m=1}^\infty r^{m-1} \begin{pmatrix} m \\ \tfrac{1}{2}(m+s) \end{pmatrix} (1-\beta)^{\tfrac{1}{2}(m+s)} \beta^{\tfrac{1}{2}(m-s)} \label{eq2}
\end{align}
This expression defines an infinite geometric series of binomial probabilities arising from all of the different possible histories by which a true hypothesis could achieve a tally of $s$, for every possible number of findings $m$. In the majority of cases, only the first few terms of the series are important, because of the leading factor $r^{m-1}$. This fact also informs us that the rate of convergence to steady state will be quite rapid, unless $r$ is large.

For any particular tally, for example $s=1$, expression~(\ref{eq2}) yields a closed-form solution like:
\begin{align}
	\hat p_{\mathrm T,1} = \frac{b(1-r)}{2 \beta r^2 }\left( \big({1 - 4 r^2 \beta(1-\beta)}\big)^{-\tfrac{1}{2}} - 1 \right)
\end{align}
For arbitrary communication parameters, the solutions have a similar structure, but are instead a series of multinomial probabilities in which the events are combinations of findings ($+$ or $-$) and communication outcomes. 

These solutions are not easy to interpret by inspection. But they do provide answers to the question: \emph{what is the probability that an hypothesis with a given tally is correct?} For any tally $s$, we can calculate:
\begin{align}
	\Pr(\text{true}|s) = \frac{\hat p_{\mathrm T,s}}{\hat p_{\mathrm T,s}+\hat p_{\mathrm F,s}}, & &
	\Pr(s|\text{true}) = \frac{\hat p_{\mathrm T,s}}{\sum_i \hat p_{\mathrm T,i}}, & &
	\Pr(s|\text{false}) = \frac{\hat p_{\mathrm F,s}}{\sum_i \hat p_{\mathrm F,i}}
\end{align}
The \emph{precision} of a tally $s$ is $\Pr(\text{true}|s)$, the proportion of hypotheses with tally $s$ that are true.  The \emph{sensitivity}, $\Pr(s|\text{true})$, is the proportion of true hypotheses with tally $s$. It indicates where the true hypotheses are. Sensitivity is important because a high precision for a tally $s$ is little help when there are few hypotheses that achieve a tally $s$. 
And the \emph{specificity}, $\Pr(s|\text{false})$, is the proportion of false hypotheses with tally $s$, indicating where the false hypotheses are. We use these definitions to explain the behavior of the system.

\subsection*{Overall dynamics}

\begin{figure}[tp]
\begin{center}
	\includegraphics[width=5.25in]{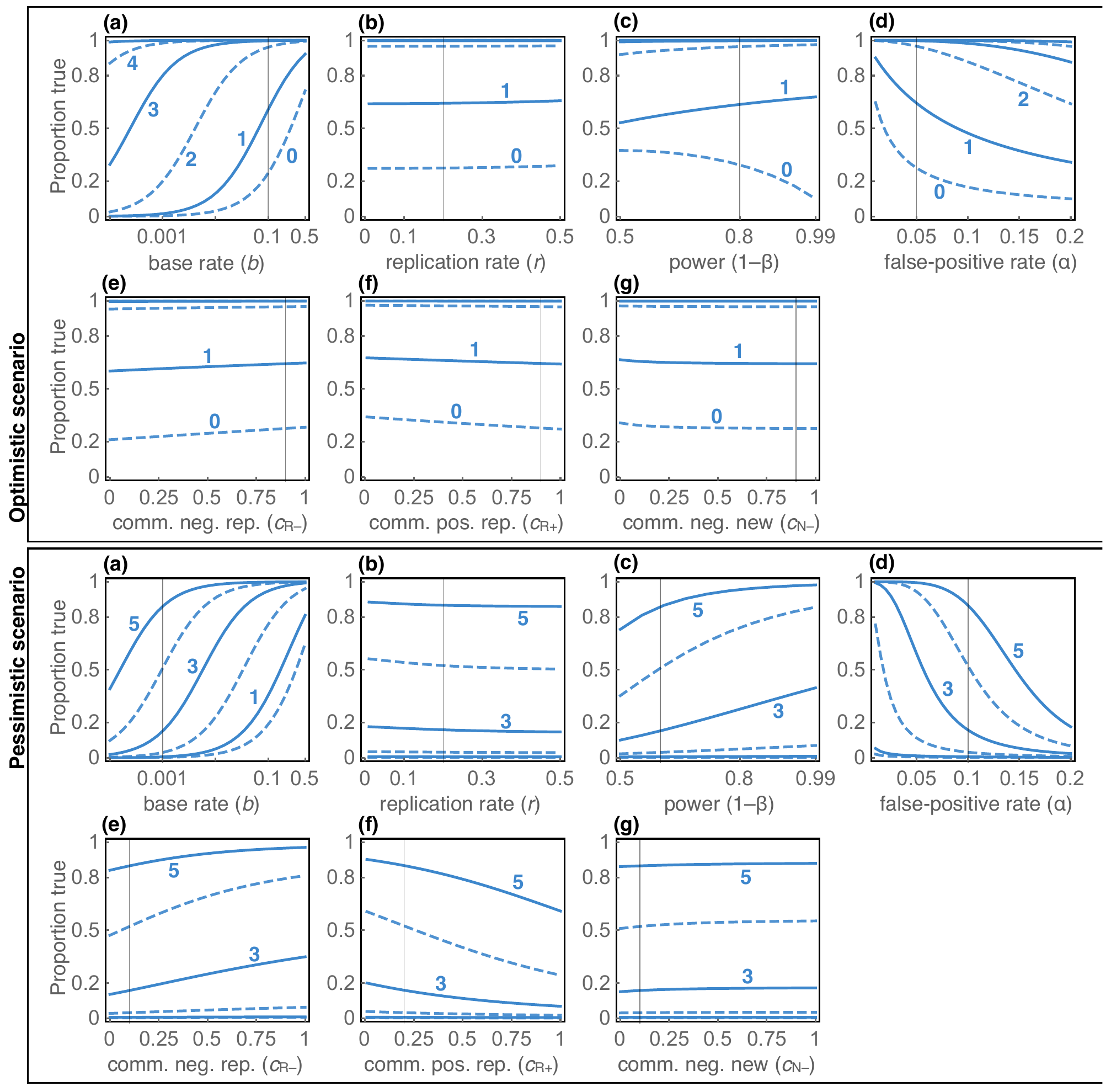}
\caption{\small Effects of base rate, replication, power, false-positives, and communication on the probability that an hypothesis with a given tally is true. The two clusters illustrate difference scenarios. The blue trends, each labeled with its tally value, show precision as it varies by the parameter on each horizontal axis.  The numbers indicate the tally of a curve. Dashed curves are tallies of an even number. The vertical hairlines show the parameter values held constant across panels within the same cluster.  }
\label{fig_precision_plots}
\end{center}
\end{figure}

Fig.~\ref{fig_precision_plots} describes the overall dynamics of precision, as a function of the different parameters. In each panel, the trend lines show the proportion of true hypotheses at each tally on the vertical axis. The tally corresponding to each trend is indicated by a number. The horizontal axis in each panel varies a single parameter. Each vertical hairline shows the value of each parameter that is held constant in other panels. This figure is complex. We'll use it to highlight the most important factors in the reliability of findings and demonstrate counter-intuitive aspects of communication. Then in the next section, we'll turn to a more general explanation of the causes of these results.

There are two clusters of plots. The top cluster represents a normatively optimistic scenario, with an auspicious base rate ($b=0.1$), unusually high power ($1-\beta=0.8$), low false-positive rate ($\alpha=0.05$), and high communication rates. The bottom cluster represents a pessimistic, or perhaps more realistic \cite{sedlemeier1989,button_power_2013}, scenario with low base rate ($b=1/1000$), lower power ($1-\beta=0.6$), higher false-positive rate ($\alpha=0.1$), and publication bias resulting in low communication of replications and negative findings. The range of base rates we show represents everything from genome wide association studies, on the low end ($b < 10^{-4}$), to predicting the winner of a presidential election, on the high end ($b=0.5$). Every scientist will have a different opinion about which values represent realism. So in the Supporting Material, we provide a Mathematica notebook for reproducing and altering these plots, so the reader can explore alternative scenarios of interest. But keep in mind that unrealistic scenarios are just as important for comprehending system dynamics.

First, notice that at tally $s=1$ very many research findings are false. In the top cluster, the base rate must get quite high before a majority of hypotheses with tally $s=1$ are true. In the bottom cluster, only the highest displayed base rates are sufficient. This dynamically replicates Ioannidis' direct calculation \cite{ioannidis_why_2005}, even in the absence of bias and multiple testing. Many initially published findings are false, unless the base rate is high, and without any invocation of fraud or researcher bias.

Second, notice that replication helps, but how much it helps varies greatly. In the top cluster, even one positive replication at $s=2$ renders most hypotheses true, at a base rate of $b=0.1$. At lower base rates, $s=3$ or $s=4$ is required to raise precision above one-half. In the bottom cluster, low power and high false-positive rate make replication quite inefficient. Even at high base rates, $s=3$ is needed. At low base rates, $s=5$ or more is required. In either cluster, achieving near-certainty that an hypothesis is true always requires replication, even with a base rate as high as $b=0.1$. In general, the same factors that make initial findings unreliable also make replications less reliable.

Note also that the rate of replication, $r$ in panel (b), has remarkably little impact. This is because replication impacts the rate at which hypotheses reach different tallies, but not so much the precision at each tally. Therefore at low replication rates, few hypotheses will ever attain $s=5$, but those that do are almost certainly true. We expand on this point in the next section. 

Third, communication of findings, panels (e-g), can both assist discovery or hinder it. Suppression of negative replications (e) reduces precision. But suppression of positive replications (f) and novel negative findings (g) either improves precision or has almost no impact on it. These aspects of the population dynamics are counter-intuitive, but quite general and revealing. The next section explains them.

\subsection*{Dynamics of communication}

The ``file drawer problem'' \cite{rosenthal_file_1979} arises when the failure to publish negative findings distorts the estimated strength of an association. We consider a related phenomenon by asking how changes in the communication parameters $c_{\mathrm N-}$, $c_{\mathrm R-}$, and $c_{\mathrm R+}$ alter the precision, sensitivity, and specificity across tallies. In the process, we'll have opportunity to explain the joint dynamics of research quality and communication biases.

In this model, it is rarely best to communicate everything. In the Supporting Material, we prove for the case of small $b$ (such that $b^2\approx0$) and small $r$ ($r^3\approx0$) that $c_{\mathrm N-} < 1$ will improve precision when $\alpha < \beta$ (usually satisfied), that $c_{\mathrm R-}<1$ improves precision when $\alpha > \tfrac{1}{2}$ (hopefully never satisfied), and that $c_{\mathrm R+} < 1$ improves precision whenever $\beta - \alpha \leq \tfrac{1}{4}$ (often satisfied). 
So some suppression of novel negative findings ($c_{\mathrm N-}<1$) and positive replications ($c_{\mathrm R+}<1$) can improve the value of replication. At larger $b$ and $r$, the conditions are more complicated, but the qualitative finding remains intact. 

To grasp why suppressing findings might help us learn what is true, think of replication as \emph{epistemological chromatography}. Chromatography is a set of techniques for separating substances that are mixed together. For example, mixed plant pigments can be separated by painting the mixture onto the tip of a strip of filter paper and then soaking the tip in a solvent. Different pigments bind more or less strongly to the solvent or the paper. Therefore as the paper absorbs the solvent, different pigments travel at different speeds, eventually separating and appearing as differently colored bands on the paper.
In the epistemological case, it is true and false hypotheses that are mixed. We wish to separate the true ones from the false.  Replication applies a ``solvent'' that  diffuses false hypotheses towards negative tallies and true hypotheses towards positive tallies. A true hypothesis diffuses upwards with probability $(1-\beta)c_{\mathrm R+}$, while a false hypothesis diffuses downwards with probability $(1-\alpha)c_{\mathrm R-}$. Thus the communication parameters adjust rates of diffusion. Just as manipulating rates of chemical diffusion can improve real chromatography, manipulating communication can improve epistemological chromatography.

\begin{figure}[tp]
\begin{center}
	\includegraphics[scale=0.75]{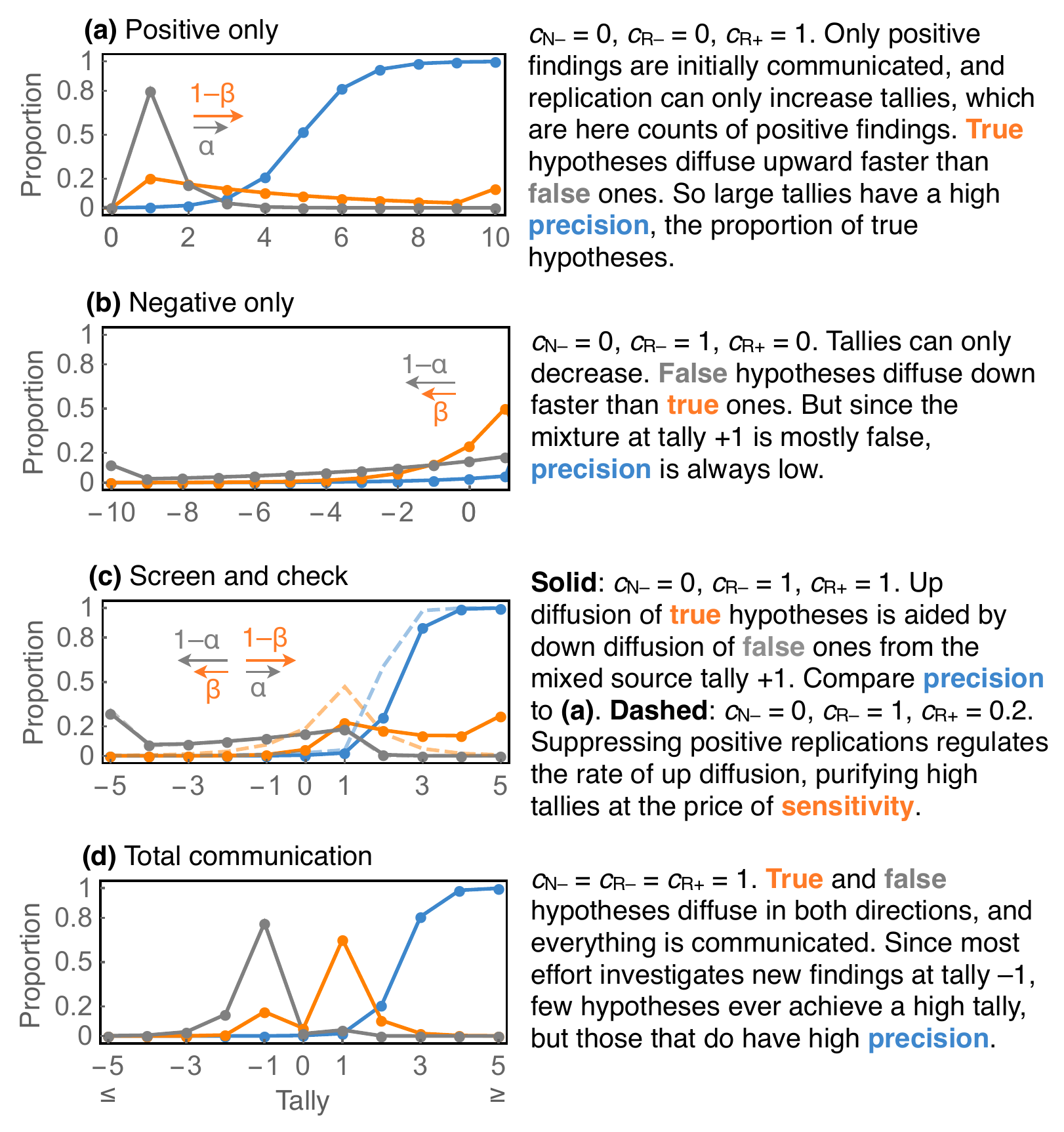}
\caption{\small Replication and communication as epistemological chromatography. Precision is indicated in blue, sensitivity in orange, and specificity in gray.}
\label{fig_communication}
\end{center}
\end{figure}

In Fig.~\ref{fig_communication}, we turn on communication one parameter at a time, in order to explain the contribution of each mode of communication to the resulting population dynamics. All four panels (a, b, c, d) show steady state precision, sensitivity, and specificity and use $b=0.001$, $r=0.2$, $1-\beta=0.8$, and $\alpha=0.05$. These values are chosen for clarity of illustration. In  the Supporting Material, we provide a Mathematica notebook to construct plots for any parameters the reader chooses. Note that for sensitivity and specificity, probability above/below the highest/lowest tally displayed is added up on the highest/lowest tally, so that none of the probability mass is hidden.

In the first three panels (a, b, c), only positive initial findings are communicated, and all new hypotheses appear at tally $s=1$. The mixture of hypotheses at this tally is heavily skewed towards false hypotheses, and so has a low precision. 
Replication may cause an hypothesis to diffuse in either direction, depending upon communication. In panel (a), negative findings are never communicated. But since true hypotheses diffuse up at a rate $1-\beta$ and false ones only at a rate $\alpha < 1-\beta$, truth is slowly separated from falsity. At tallies of 8 or more, nearly all hypotheses are true, as indicated by the precision. Note however that most true hypotheses that have been communicated at all exist at low tallies, as indicated by the sensitivity. With enough time and replication, every true hypothesis can be split from the false. 
This is unlike the case in panel (b), where only negative replications are communicated. The same dynamic works in reverse here, and replication creates a pure  sample of false hypotheses at low tallies. 

Combining both directions of diffusion is synergistic, as illustrated in panel (c). 
Now both positive and negative replications are communicated. The downward diffusion of false hypotheses makes the upward diffusion of true hypotheses more efficient. This effect arises because $1-\alpha > 1-\beta$. False hypotheses diffuse down faster than true hypotheses diffuse up. This purifies the source mixture at $s=1$, allowing for precision to approach high values at much smaller tallies than in the absence of either diffusion process. In this example, hypotheses with tallies of $s=3$ and greater are true more than 80\% of the time, and the sensitivity indicates that more than half of all published true hypotheses have a tally of 3 or more. Keep in mind that this 80\% is equally interpretable as a probability that applies to a unique hypothesis. So it provides epistemic value, independent of the frequency interpretation.

Diffusion in both directions is enhanced by suppressing some positive replications. The dashed curves in panel (c) provide a comparison when only 20\% of positive replications are communicated. Precision is substantially higher in this case, but at the cost of reduced sensitivity at high tallies. This effect arises from the same dynamic as before: by setting $c_{\mathrm R+}<1$, we have effectively slowed all upward diffusion. This allows rapid downward diffusion from negative replications to further clean the source mixture, but at the cost of diffusing more true hypotheses towards negative tallies. This dynamic is beneficial when base rate is especially low. So we achieve a very clean sample of truth at smaller positive tallies in this scenario, but at the price of finding fewer true hypotheses in total. Whether this is an improvement depends upon context, an issue we take up in the discussion.

Finally, full communication is illustrated in panel (d). High precision is achieved at high tallies, but few hypotheses reside at those tallies. This inefficiency arises from the unbiased allocation of replication effort. When all initial findings are communicated, replication effort is overwhelmed by following up on initial negative findings, the spike in specificity seen at tally $s=-1$. When the base rate is low, it can be better to screen for positive findings than to publish every negative finding. 
Note however that increasing precision, the proportion of hypotheses at a given tally that are true, is not necessarily the only objective. It does us little good if sensitivity is very low at all high tally values. We return to this point in a later section, when we consider differential power and false-positive rates between initial studies and replications.

\subsection*{Targeted replication}

Replication in the preceding analysis is purely random: every communicated hypothesis has an equal chance of being the target of a replication effort. Targeting particular tally values, like $s=1$, might be more efficient. Here, we demonstrate that the main effect of targeted replication is to improve sensitivity, the proportion of true hypotheses at positive tallies. It has little effect on precision, the proportion of hypotheses at positive tallies that are true.

To modify the population dynamics to allow targeted replication effort, assume that a proportion $r_{\mathrm T}$ of all replication attempts target a chosen list of tally values, selecting an hypothesis randomly from all hypotheses within the list. For example, this list might consist of all previously communicated hypotheses with a positive tally of three or less, so that researchers concentrate their replication efforts on hypotheses thought to be true but with relatively high uncertainty. The rest of the time, $1-r_{\mathrm T}$, replication effort remains unbiased. 

\begin{figure}[tp]
\begin{center}
	\includegraphics[scale=0.75]{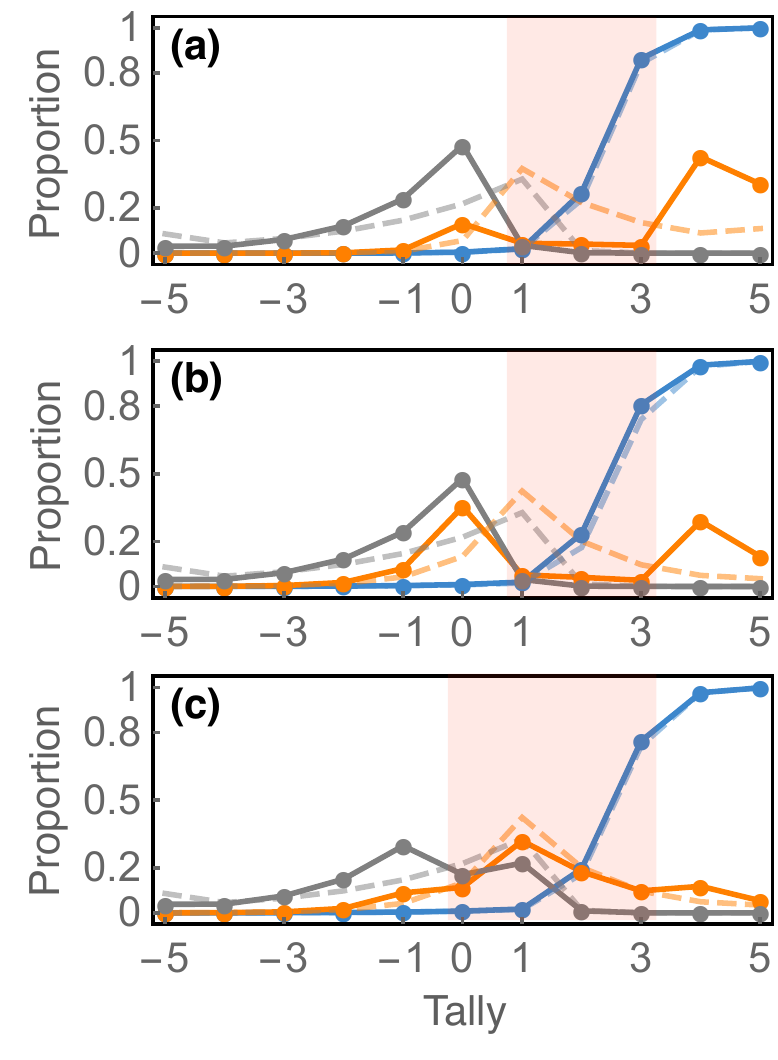}
\caption{\small Targeted replication effort. In all three plots, tallies marked for targeted replication are shown by the shaded region. Precision is indicated in blue, sensitivity in orange, and specificity in gray. Baseline parameters set to $b=0.001$, $\alpha=0.05$, $r=0.1$, $r_{\mathrm T}=0.5$, $c_{{\mathrm N}-}=0$, $c_{{\mathrm R}-}=c_{{\mathrm R}+}=1$. Dashed curves display steady-state without targeted replication, $r_{\mathrm T}=0$. (a) High power setting, $1-\beta=0.8$. (b) Low power setting, $1-\beta=0.6$. (c) Low power, $1-\beta=0.6$, and including tally $s=0$ in the target. }
\label{fig_target}
\end{center}
\end{figure}

Fig.~\ref{fig_target} shows the resulting modification of the dynamics. The dashed curves in these plots show the steady-state dynamics in the absence of targeting. The shaded pink regions show the range of tally values included in the target. In each case, targeting improves sensitivity at higher positive tallies. Thus it helps to diffuse true hypotheses towards tallies with very high precision. But there is very little effect on precision itself. Targeting helps because it directs effort towards tallies that may not have a high density of hypotheses. When replication effort is unbiased, most effort is directed to tallies where the bulk of hypotheses reside. Therefore when the target range includes a wide range, as in panel (c), it becomes relatively ineffective. 

Why doesn't targeting improve the proportion of hypotheses that are true at higher tallies? Targeting serves mainly to speed up diffusion, without altering the {\em relative} rates at which true and false hypotheses diffuse. Changes in communication rates, in contrast, do alter the differential rates of diffusion, and so may dramatically alter precision, as seen in the previous section.

\subsection*{Differential power and false-positives}

So far, we have assumed that power $1-\beta$ and false-positive rate $\alpha$ are the same in initial studies and replications. Differences between initial studies and replications have been at the center of concerns about replication \cite{bissell_reproducibility_2013}. Here we analyze a version of our model in which we allow the power and false-positive rate to vary. Let $1-\beta_{\mathrm R}$ and $\alpha_{\mathrm R}$ be the power and false-positive rate, respectively, for replications. What effects do both higher-powered replication and lower-powered replication have on dynamics? 

\begin{figure}[tp]
\begin{center}
	\includegraphics[scale=0.75]{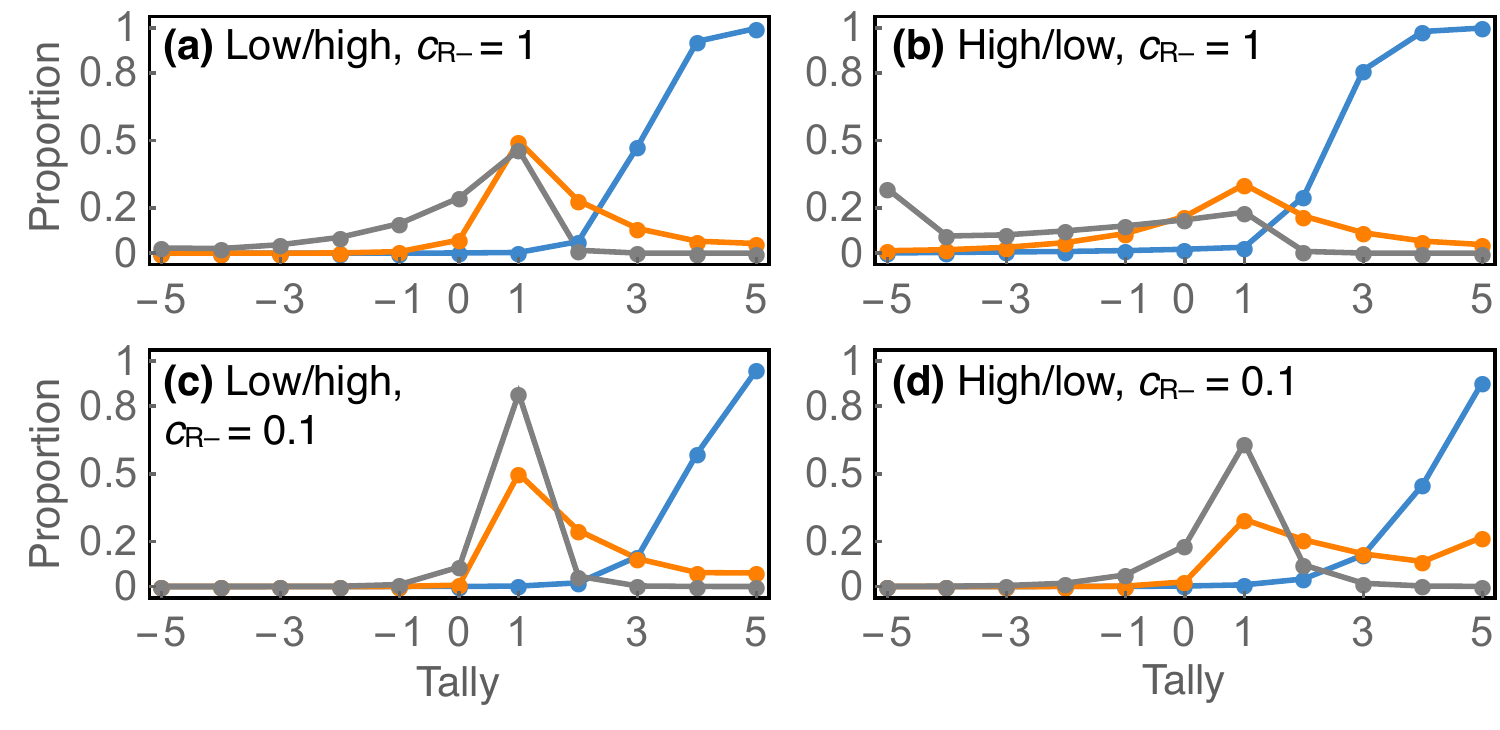}
\caption{\small Differential power and replication dynamics. Precision is indicated in blue, sensitivity in orange, and specificity in gray. (a) Low power initial studies ($1-\beta=0.6$, $\alpha=0.2$) but high power replications ($1-\beta_{\mathrm R}=0.8$, $\alpha_{\mathrm R}=0.05$). (b) High power initial studies ($1-\beta=0.8$, $\alpha=0.05$) but low power replications ($1-\beta_{\mathrm R}=0.5$, $\alpha_{\mathrm R}=0.05$). (c) and (d)  as in (a) and (b), respectively, but only 10\% of negative replications are communicated.}
\label{fig_power}
\end{center}
\end{figure}

In Fig.~\ref{fig_power}, we present two extreme, illustrative scenarios. Both scenarios use $b=0.001$, $c_{{\mathrm N}-}=0$, $c_{{\mathrm R}-}=c_{{\mathrm R}+}=1$, $r=0.2$, and $r_{\mathrm T}=0$ unless noted otherwise.
The first is a ``low/high'' scenario in which initial findings are produced by studies with $1-\beta=0.6$ and $\alpha=0.2$, but replications have conventional $1-\beta_{\mathrm R}=0.8$ and $\alpha_{\mathrm R}=0.05$. This scenario reflects a context in which initial studies use small samples and suffer from motivated data-snooping or data-contingent analysis that elevates false-positives \cite{gelman_ethics_2014,simmons_false-positive_2011-1}. This scenario is shown in panel (a).
The second scenario is a ``high/low'' scenario, with $1-\beta=0.8$, $\alpha=0.05$, $1-\beta_{\mathrm R}=0.5$, $\alpha_{\mathrm R}=0.05$. This scenario reflects a context in which replications are prone to error, because a true effect requires skill to produce \cite{bissell_reproducibility_2013}. This scenario is shown in panel (b).

Comparing the two, notice that low/high is more damaging overall, as the elevated false-positives cascade through the population during diffusion of hypotheses to higher tallies. Thus it takes more replication in (a) to achieve the same precision as in the high/low scenario (b). Even with only 50\% power in (b), replication successfully separates true hypotheses from false ones. Unfortunately, it also diffuses many true hypotheses towards negative tallies. The high precision at positive tallies is a result of a false hypothesis' relative inability to attain a positive replication, not a result of a true hypothesis' ability to avoid a negative replication.

In the last two panels, (c) and (d), we show how these scenarios change when negative replications are suppressed, $c_{{\mathrm R}-}=0.1$. The situation generally worsens in both cases, but failure to communicate negative replications does prevent true hypotheses from attaining negative tallies, in the case in which replication power is low, (d).

Overall, replications continue to have value, even when they are more prone to error than original studies. As long as true hypotheses are more likely to diffuse upwards than downwards, replication aids discovery.

\section*{Discussion}
Ours is the first analytical model of the joint population dynamics of scientific hypothesis generation, communication, and replication. 
Such a model is necessary to illuminate debates about scientific practice, because until researchers report the results of every study, empirical estimates of base rate are  not possible. And without consideration of population dynamics, any discussion of the value of research findings remains at least partly na\"{i}ve, because it is notoriously difficult to reason verbally about complex systems. Our model produces a number of valuable counter-intuitive results. But even when its results are intuitive, some model like ours is needed to demonstrate their logic. It is not enough to merely hold the correct belief; we must also justify that belief.

This model is not a definitive representation of the scientific process, nor does it aim to be. It omits many relevant factors, such as investigator bias and disagreements about the interpretation of evidence. These omissions allow the model to address focused questions about the evidential value of research as it emerges from the joint dynamics of hypothesis generation, replication, and communication. Models that account for more and different factors must also include variants of these complex dynamics, so our model is a necessary and useful first step.

Our analysis re-emphasizes what every textbook says: replication is an essential aspect of scientific discovery. However, it also quantifies its impact and emphasizes that replication itself can be unreliable---the factors that make initial findings unreliable also make replication less reliable. When base rate is low, power is low, or false positives common, then many successful replications will be needed to attain confidence in an hypothesis. This is especially true when negative replications are difficult to publish. 

We find that low base rate and high false positive rate are the most important threats to the effectiveness of research, replicated or not. This re-emphasizes the importance of quality theorizing, in order to improve base rate. While it is appealing to think that science works regardless of where hypotheses come from, undisciplined hypothesis generation reduces base rate and makes initial findings mostly false. Then  large amounts of replication will be needed to uncover the truth. In fields such as physics and evolutionary biology, a great deal can be and is done to vet theory in the realm of pure thought, using mathematics and simulation. But in fields such as social psychology, theory development is rarely formalized \cite{smaldino_abm_2015}.

The results also re-emphasize the value of efforts to suppress false positive findings, such as pre-registered data analysis plans. It is important to recognize that any single scientific hypothesis may correspond to many different statistical hypotheses. If a statistical hypothesis can be chosen after seeing the data,  reasonable scientific hypotheses can become unreasonably flexible \cite{Gelman:2013aa}. And many data-contingent transformations and modeling choices that increase power, conditional on an hypothesis being true, will also increase false-positives, conditional on the hypothesis being false. For example, dropping outliers may well aid discovery, if the hypothesis is true. But it may also dramatically inflate false-positives, if the hypothesis is not true \cite{BakkerOutliers}.

Our model immediately informs debates over the meaning of failed replications. For example, some have suggested that positive replications have more worth than negative replications \cite{schnall_clean_2014}, or even that failed replications ``cannot contribute to a cumulative understanding of scientific phenomena'' \cite{mitchell_emptiness_2014}. We find the opposite: communicating a failure to replicate is typically more informative than communicating a successful replication. This remains true even when replication attempts have lower power than original studies. However, a single failure to replicate is entirely consistent with a true hypothesis in many scenarios. So both positive and negative replications may be regarded with skepticism. But neither is without value. Of course our model is merely a model. But unlike the verbal arguments we cite, it is at least clear in its assumptions, and its logic can be verified.

Our model also sheds light on proposals for improving the reliability of research. For example, many have called for pre-registration and review with a commitment from journals to publish research results, positive or negative, in order to reduce under-reporting of negative findings \cite{Nyhan2014}. Our analysis suggests that these proposals should distinguish between new hypotheses and replication attempts. If indeed many new hypotheses are false in many fields, a pre-registration process would merely fill journal pages with null findings, doing great harm by crowding out candidate hypotheses that have passed an initial screening. In our model, there is little harm in ignoring novel negative findings, because they add very little information. 
Indeed, Figure 2 illustrates that the effect of ignoring novel negative results on precision is negligible.
In contrast, a negative replication may add a lot of information. We suspect however that our model exaggerates this effect, because the model ignores the wasted effort arising from different researchers repeating an investigation in ignorance of one another's negative findings. 
And there are certainly fields in which full publication may be the best policy, such as when false-positive rates are low or when the total number of testable hypotheses is very small. 
Nevertheless, the qualitative difference in information value between novel and follow-up negative findings will remain as long as the base rate in the published literature is higher than it is in novel investigations.

The model stimulates empirical investigation by clarifying which factors must be estimated in order to gauge the evidential value of research, as well as being readily translatable into a statistical framework, due to its analytical specification. Our model provides an implicit `null model' of research: setting $b=0$ provides a null distribution of novel findings and lifespans of hypotheses. Null models are deliberately unrealistic and usually \emph{a priori} false, but have nevertheless played an important role in science \cite{Wimsatt:1987aa}. 

There are additional factors to address in future work. Our model ignores researcher bias, multiple testing, and data snooping, each of which deflates base rate or inflates false-positive rate. Our analysis is framed in a standard, but unsatisfying, ``true'' and ``false'' classification, rather than considering practical significance and effect size estimation \cite{gelman_ethics_2014}. Our model can be directly generalized to consider variation in effect size instead of true and false hypotheses. We explain this generalization in  the Supporting Material. However, our model does not directly address causal inference nor point estimation. 

Incentives also matter. A dynamic analysis of strategic behavior under different incentive structures would aid policy analysis \cite{Kitcher2000}. As Karl Popper argued, science does not work because scientists are selfless and unbiased people.  Rather it works because its institutions channel our bias into the production of public goods \cite{PopperMythFramework}. In particular, we worry that a research environment that lacks replication may actually select for statistical practices that inflate false-positives, as labs with such practices can more readily publish findings and place students in new positions, all while outrunning the truth.

Replication may offer other benefits that are not accounted for in our model. A failed replication may be valuable because it inspires a new hypothesis in order to explain variation in findings. When findings do not generalize across samples, this creates an opportunity to explain the variation \cite{henrichetalScience2010,Scott07102014}. In our view, the goal of replication is not merely to find the same result, but also to discover how a result arises and how it is likely to vary in realistic, non-laboratory, contexts.

Despite these shortcomings, our model provides specific quantitative evaluations of many verbal arguments, as well as drawing attention to the population dynamics of scientific knowledge. Science is a subtle project. Understanding it demands the same rigor that we apply to projects within science itself.

\newpage

\begin{center}
 {\bf
{\large SUPPLEMENTAL INFORMATION\\~\large Replication, Communication, and the Population Dynamics of Scientific Discovery}
}
\end{center}

\section{Derivation of full model with random replication}

Let $f_{\mr{T},s} = n_{\mr{T},s}/n$ be the frequency of true hypotheses with tally $s$. Under the assumptions and definitions supplied in the main text, the full recursion for $n_{\text{T},s}^\prime$ is given by:
\begin{align}
	n_{\mr{T},s}^\prime &= n_{\mr{T},s} + anr 
		\big( 
			-f_{\mr{T},s}( c_{\mr{R}+}(1-\beta) + c_{\mr{R}-}\beta ) + 
			f_{\mr{T},s-1}(1-\beta) c_{\mr{R}+} + 
			f_{\mr{T},s+1}\beta c_{\mr{R}-}
		\big)
\end{align}
for $s$ not equal to 1 or $-1$. In those cases, there is an additional term. For $s=1$:
\begin{align}
	n_{\mr{T},1}^\prime &= n_{\mr{T},1}  \\
		&\quad + anr 
		\big( 
			-f_{\mr{T},1}( c_{\mr{R}+}(1-\beta) + c_{\mr{R}-}\beta ) + 
			f_{\mr{T},0}(1-\beta) c_{\mr{R}+} + 
			f_{\mr{T},2}\beta c_{\mr{R}-}
		\big) \nonumber \\
		& \quad + an(1-r)b(1-\beta) \nonumber
\end{align}
The $an(1-r)b(1-\beta)$ term accounts for inflow of novel positive findings, all of which are communicated. For $s=-1$:
\begin{align}
	n_{\mr{T},-1}^\prime &= n_{\mr{T},-1}  
	\\ &\quad + anr 
		\big( 
			-f_{\mr{T},-1}( c_{\mr{R}+}(1-\beta) + c_{\mr{R}-}\beta ) + 
			f_{\mr{T},-2}(1-\beta) c_{\mr{R}+} + 
			f_{\mr{T},0}\beta c_{\mr{R}-}
		\big) \nonumber \\
		& \quad + an(1-r)b\beta c_{\mr{N}-} \nonumber
\end{align}
The $an(1-r)b\beta c_{\mr{N}-}$ term accounts for inflow of novel negative findings, only $c_{\mr{N}-}$ of which are communicated. 
Recursions for false hypotheses can be derived just by substitution of variables: $b \rightarrow 1-b$ and $1-\beta \rightarrow \alpha$. 

These recursions implicitly define the population growth recursion for $n$:
\begin{align}
	n^\prime &= n + an(1-r)
		\big(
			b( 1-\beta + \beta c_{\mr{N}-} )
			+ (1-b)( \alpha + (1-\alpha) c_{\mr{N}-} )
		\big)
\end{align}
This just indicates that the population of published hypotheses grows proportional to the innovation rate, $1-r$, and the rates at which true and false hypotheses respectively produce positive and negative findings, as well as the rate at which negative findings are communicated.

\section{Beyond ``true'' and ``false''}

Above we noted that recursions for false hypotheses can be derived just by substitution of variables: $b \rightarrow 1-b$ and $1-\beta \rightarrow \alpha$. In other words, true and false hypotheses are differentiated only by the rate at which they appear in new investigations and their respective probabilities of producing positive findings. This also means it is straightforward to expand the model to additional epistemic states, as ``true'' and ``false'' really just more more and less correct. For example, small, medium, and large effect sizes could be represented by three states, each with its own base rate and probability of producing a positive result. The derivation would remain the same, but an additional set of steady-state solutions would appear.

\section{Steady-state solutions}

We have analyzed this model using a variety of methods. First, we solved the model analytically for every structure except for targeted replication (to be defined later). Second, when analytical solution was not possible, we solved the model numerically. Third, we studied the model under both deterministic and stochastic simulations, written independently by both authors in different programming languages. All forms of analysis yield identical results.

The model above can be solved directly, in one of two ways. First, it can be solved exactly by bounding tallies within a minimum and maximum (using either absorbing or reflecting boundaries) and then solving the system of simultaneous equations for values of the state variables $f_{i,s}$ for $i \in \{\mr{T},\mr{F}\}$. This approach is probably the most straightforward. Second, it can be solved to any level of approximation desired by iteratively solving the system of equations outward from $s=0$.

Both approaches yield solutions that take the form of closures of infinite geometric series expressions. Using these solutions, we found the unbounded infinite series solution based upon intuition---\emph{ansatz} is what our mathematics instructors used to call it. Since the solutions from the brute-force approach looked like closures of infinite series, and the simulation results produced what resembled a mixture of geometric series, we guessed the underlying limiting distribution. We then verified our \emph{ansatz} solution by plugging it back into the recursions and also by comparing it to numerical results and our previous solutions. Finally, we induced the infinite series representation by constructing Taylor series expansions of the closed series expressions, yielding the sequential terms of the solution expression in the next section.

\subsection{Full communication solution}

Here we repeat the simplest such solution from the main text and then motivate its justification. The steady state proportion of hypotheses that are both true and have tally $s$, when all findings are communicated, is given by:
\begin{align}
	\hat p_{\mathrm T,s} = b(1-r) \sum_{m=1}^\infty r^{m-1} K \big( m , (m+s)/2 \big) (1-\beta)^{\tfrac{1}{2}(m+s)} \beta^{\tfrac{1}{2}(m-s)}
\end{align}
where $K(m,(m+s)/2)$ is the number of ways to get $(m+s)/2$ positive findings in $m$ investigations of the same hypothesis. This is simple the binomial chooser, but implicitly evaluating to zero whenever $(m+s)/2$ is not an integer. Since $s$ is the difference between the number of positive and negative findings, this multiplicity accounts for the number of paths by which an hypothesis can be studied $m$ times and end up with a tally $s$. The remaining terms leading with $1-\beta$ and $\beta$ are just the probabilities of getting $(m+s)/2$ positive findings and $(m-s)/2$ negative findings, respectively.

Here's how to motivate the above solution. For any given tally $s$, there are an infinite number of histories by which it could have ended up with that tally. 
\begin{itemize}

\item Consider tally $s=1$, for example. If the hypothesis is true, it could end up most simply at $s=1$ with just one initial positive finding. This happens with probability $(1-r)b(1-\beta)$, indicating innovation times base rate of true hypotheses times the probability of an initial positive finding.

\item Similarly, if instead the hypothesis has been studied twice, which happens $(1-r)br$ of the time, the number of ways it could end up with $s=1$ is exactly zero, and the multiplicity handles this by assigning $K(2,(2+1)/2)=0$. 

\item For three studies, there are $K(3,2)=3$ ways $s=1$ could happen. Represented as sequences of positive and negative findings, these are: (1) $++-$, (2) $+-+$, and (3) $-++$. The probability of any one of these is $(1-\beta)^2 \beta$, and the probability that an hypothesis is true and has been studied three times is $(1-r)br^2$.
\end{itemize}
The pattern here generalizes so that the total probability is just:
\begin{itemize}

\item the sum over number of studies on an hypothesis from $m=1$ to $m=\infty$ of the probability the hypothesis was studied $m$ times, given by $(1-r)r^{m-1}$

\item times the number of ways it could end up with a tally $s$ in $m$ steps, given by $K(m,(m+s)/2)$

\item times the probability of getting $(m+s)/2$ positive and $(m-s)/2$ negative findings.

\end{itemize}
Writing down this summation and factoring out the common term $b(1-r)$ completes the expression.

This steady-state solution obviously assumes that there has been an infinite amount of research time, such that every $m$ can be realized. In practice, since the sequence is geometric in $r$, the probabilities of higher values of $m$ decline very rapidly and simulations confirm that steady-state is reached quite rapidly, as long as the replication rate $r$ is not close to $r=1$. 

More importantly we think, these solutions are never meant to describe actual science, but rather to allow us to reason about causal forces in actual science. So the steady state expressions are important even if, as in many real dynamical system, they are never exactly realized. For example, problems in evolutionary theory are routinely solved by asking what happens on the infinite time horizon. Such solutions have been incredibly useful, despite the fact that no real population or environment is stationary enough to make the exercise literally sensible. 

\subsection{Arbitrary communication solution}

When communication parameters are allowed to be less than one, the above strategy generalizes directly, but does become complex. The expressions get much more complex, because now the infinite series is over multinomial probabilities of three possible outcomes at each replication investigation of an hypothesis: (1) positive and communicated, (2) negative and communicated, or (3) not communicated. In addition, when findings are not always communicated, then the effective activity rate changes, making other probabilities conditional on observable activity. Still, these solutions can be derived both by the logic to follow or by brute-force solution of the system of recursions. Solving the system of recursions does allow for easily defining reflecting or absorbing tally boundaries, which may be appealing in some contexts. The combinatoric solution to follow assumes unbounded tallies. Solutions in the bounded and unbounded cases are nearly identical, for all scenarios considered in the main text. The Mathematica notebooks in the supplemental materials present code for both types of solution.

We present the solutions here as a sequence of conditional probabilities, as we've found this form easier to interpret than the general multinomial form. Therefore they provide more insight. Specifically, we decompose the multinomial probabilities into a binomial series for observed/unobserved investigations of a hypothesis and a binomial series for positive/negative findings conditional on being observed. The solutions take the form:
\begin{align}
	\hat p_{\mr T,s} &= \Pr(\mr T) \Pr( \text{activity} ) \Pr( \text{new} | \text{activity} ) \big( (1-\beta) \Pr(s|+) + \beta c_{\mr{N}-} \Pr(s|-) \big)
\end{align}
Where:
\begin{align}
	\Pr(\mr T) &= b\\
	\Pr(\text{activity}) &= r + (1-r) \big( b( (1-\beta) + \beta c_{\mr{N}-} ) + (1-b)( \alpha + (1-\alpha)c_{\mr{N}-} ) \big)\\
	\Pr( \text{new} | \text{activity} ) &= \frac{(1-r) \big( b( (1-\beta) + \beta c_{\mr{N}-} ) + (1-b)( \alpha + (1-\alpha)c_{\mr{N}-} ) \big)}{\Pr(\text{activity})}
\end{align}
The probabilities $\Pr(s|+)$ and $\Pr(s|-)$ give the probabilities of tally $s$ averaging over number of investigations $m$ and un-communicated findings $u$, beginning with either a positive finding or a negative finding, respectively. This conditioning is necessary because a tally $s$ can be reached by different paths once communication is partial. These probabilities are given by:
\begin{align}
	\Pr(s|+) &= I_1(s) + \sum_{m=1}^{\infty} \sum_{u=0}^m R^m \Pr(u|m) S(s-1|m-u)\\
	\Pr(s|-) &= I_{-1}(s) + \sum_{m=1}^{\infty} \sum_{u=0}^m R^m \Pr(u|m) S(s+1|m-u)
\end{align}
where $I_a(b)$ is a function that returns 1 when $a=b$ and zero otherwise and $R=r/\Pr(\text{activity})$ is the probability of replication, conditional on activity as defined earlier. The term $\Pr(u|m)$ gives the probability of $u$ un-communicated findings in $m$ investigations, defined as:
\begin{align}
	\Pr(u|m) &= \frac{m!}{u!(m-u)!} q_\circ^u (1-q_\circ)^{m-u}
\end{align}
where
\begin{align}
	q_\circ &= (1-\beta_{\mr R}) (1-c_{\mr{R}+}) + \beta_{\mr R} (1-c_{\mr{R}-})
\end{align}
is the probability a replication finding is un-communicated, averaging over positive and negative findings. Finally, the function $S(z|n)$ provides the probability that a sequence of length $n$ communicated replication findings producing a difference $z$ between positive and negative replications. It is defined as:
\begin{align}
	S(z|n) &= 
		\begin{cases}
			I_0(z) & \text{if $n = 0$}\\
			K( n , (n+z)/2 ) q_+^{(n+z)/2} (1-q_+)^{(n-z)/2} & \text{if $n>0$}
		\end{cases}
\end{align}
where $K(a,b)$ is again the binomial chooser function, but evaluating to zero when $b$ is not an integer, and:
\begin{align}
	q_+ &= \frac{(1-\beta_{\mr R})c_{\mr{R}+}}{1-q_\circ}
\end{align}
which is the probability of a positive replication, conditional on the replication finding being communicated.

\section{Approximate conditions for reduced communication}

We argue in the main text that full communication is rarely optimal, from the perspective of precision. Consider the full communication context: $c_{\mr{N}-}=c_{\mr{R}-}=c_{\mr{R}+}=1$. For small $b$ ($b^2 \approx 0$) and small $r$ ($r^3 \approx 0$), precision as defined in the main text is improved by reducing communication parameters under the following conditions:
\begin{itemize}
\item $c_{\mr{N}-}<1$ when $\alpha < \beta$ (easy to satisfy)
\item $c_{\mr{R}-}<1$ when $\alpha > 0.5$ (hopefully not satisfied)
\item $c_{\mr{R}+}<1$ when $\beta - \alpha \leq 1/4$
\end{itemize}
These conditions are derived by first defining precision at $s=1$, which is most conservative precision to investigate, because it benefits the least from replication, and higher tallies always have higher precision than $s=1$. So improvements at $s=1$ cascade upwards to higher tallies. Let $\text{PPV}_1$ be the precision at $s=1$. Then the first condition is proved by computing the derivative $\partial \text{PPV}_1 / \partial c_{\mr{N}-}$, evaluated at full communication parameter values. Then Taylor expand the result simultaneously by second-order around $r=0$ and by first-order around $b=0$. Neglecting terms of order $O(b^2)$ and $O(r^3)$ and higher:
\begin{align}
	\frac{\partial \text{PPV}_1}{\partial c_{\mr{N}-}} \approx 
				- r^2 \frac{1-\beta}{\alpha} b (\beta-\alpha)(1-\beta-\alpha)(5-6\alpha)
\end{align}
which is negative unless $\alpha > \beta$. Thus suppressing some initial negative findings is favorable, provided the base rate is small and replication is not too common. We think most scientific fields satisfy these conditions, but reasonable people can and do disagree on that point.

In contrast, suppressing negative replications is unlikely to help. By the same strategy, but this time differentiating with respect to $c_{\mr{R}-}$:
\begin{align}
	\frac{\partial \text{PPV}_1}{\partial c_{\mr{R}-}} \approx 
				r b \frac{1-\beta}{\alpha} (1-\beta-\alpha) (1 + 2r(\beta-\alpha) )
\end{align}
which is guaranteed positive, indicating that $c_{\mr{R}-} = 1$ is favored, when $\alpha \leq 0.5$, because by assumption $1-\beta > \alpha$.

The third condition is derived similarly:
\begin{align}
	\frac{\partial \text{PPV}_1}{\partial c_{\mr{R}+}} \approx 
				- b r \frac{1-\beta}{\alpha} (1-\beta-\alpha) ( 1 - 4r(\beta-\alpha) )
\end{align}
The last term is the one in play. For the above to be negative, it is required that:
\begin{align}
	r < \frac{1}{4}\frac{1}{\beta-\alpha}
\end{align}
And this is guaranteed when $\beta-\alpha \leq 1/4$.

\clearpage
\bibliographystyle{plos2015} 
\bibliography{bofflandia}

\end{document}